\documentclass[aps,prb,twocolumn,floatfix,amsmath,amssymb,showpacs,
               superscriptaddress]{revtex4-1}
\usepackage{graphicx,placeins}
\usepackage{color}
\usepackage{dcolumn}
\usepackage{bm}
\usepackage{epsfig,psfrag,amsmath,amssymb,float}
\input{epsf}
\usepackage{hyperref}
\usepackage[percent]{overpic}
\hypersetup{
    colorlinks=true,
    linkcolor=blue,
    citecolor=blue,
    filecolor=magenta,      
    urlcolor=cyan,
}
\usepackage{array}
\newcolumntype{C}[1]{>{\centering\arraybackslash$}p{#1}<{$}}

\setcitestyle{square}
\newcommand{\pdag}{{\phantom{\dagger}}}
\bibpunct{[}{]}{,}{n}{}{}
\begin{document}
\title{Novel Block Excitonic Condensate at $n=3.5$ in a \\
Spin-Orbit Coupled $t_{2g}$ Multiorbital Hubbard Model}

\author{Nitin Kaushal}
\affiliation{Department of Physics and Astronomy, The University of 
Tennessee, Knoxville, Tennessee 37996, USA}
\affiliation{Materials Science and Technology Division, Oak Ridge National 
Laboratory, Oak Ridge, Tennessee 37831, USA}
\author{Alberto Nocera}
\affiliation{Department of Physics and Astronomy, The University of 
Tennessee, Knoxville, Tennessee 37996, USA}
\affiliation{Materials Science and Technology Division, Oak Ridge National 
Laboratory, Oak Ridge, Tennessee 37831, USA}
\affiliation{Quantum Matter Institute, University of British Columbia, Vancouver, British Columbia V6T 1Z4, Canada}
\author{Gonzalo Alvarez}
\affiliation{Center for Nanophase Materials Sciences, Oak Ridge National Laboratory, Oak Ridge, Tennessee 37831, USA}
\affiliation{Computational Science and Engineering Division, Oak Ridge National Laboratory, Oak Ridge, Tennessee 37831, USA}
\author{Adriana Moreo}
\affiliation{Department of Physics and Astronomy, The University of 
Tennessee, Knoxville, Tennessee 37996, USA}
\affiliation{Materials Science and Technology Division, Oak Ridge National 
Laboratory, Oak Ridge, Tennessee 37831, USA}
\author{Elbio Dagotto}
\affiliation{Department of Physics and Astronomy, The University of 
Tennessee, Knoxville, Tennessee 37996, USA}
\affiliation{Materials Science and Technology Division, Oak Ridge National 
Laboratory, Oak Ridge, Tennessee 37831, USA}

\date{\today}

\begin{abstract}
Theoretical studies recently predicted the condensation of spin-orbit excitons 
at momentum $q$=$\pi$ in $t_{2g}^4$ spin-orbit coupled three-orbital Hubbard models at electronic density $n=4$.
In parallel, experiments involving iridates with non-integer valence states for the Ir ions are starting to 
attract considerable attention. 
In this publication, using the density matrix renormalization group 
technique we present evidence for the existence of a novel excitonic condensate 
at $n=3.5$ in a one-dimensional 
Hubbard model with a degenerate $t_{2g}$ sector, when
in the presence of spin-orbit coupling. At intermediate Hubbard $U$ 
and spin-orbit $\lambda$ couplings, we found an excitonic condensate at the unexpected momentum 
$q$=$\pi/2$ involving $j_{\textrm{eff}}=3/2,m=\pm1/2$  and $j_{\textrm{eff}}=1/2,m=\pm1/2$ bands 
in the triplet channel, 
coexisting with an also unexpected block magnetic order. We also present the entire 
$\lambda$ vs $U$ phase diagram, at a fixed and robust Hund coupling. 
Interestingly, this new ``block excitonic phase'' is present even at 
large values of $\lambda$, unlike the $n=4$ excitonic phase discussed before. Our computational 
study helps to understand and predict the possible magnetic phases of materials with $d^{3.5}$ valence 
and robust spin-orbit coupling.

\end{abstract}
\maketitle

\section{Introduction}

In the last decade, the $4d$/$5d$ transition metal oxides have received considerable 
attention in the Condensed Mater community, specially because they
provide a unique platform for the development of unconventional magnetic and transport properties 
mainly as a consequence of a robust spin-orbit coupling \cite{Cao03,ir1,ir2,ir3,ir4,ir5,ir6,ir7,ir8,ir9,ir10}. 
One of the most interesting materials 
is Sr$_{2}$IrO$_{4}$ containing Ir$^{4+}$ ions, with an electronic density $n=5$ \cite{BJKim01}. 
This compound displays similarities with La$_{2}$CuO$_{4}$, even with a relatively smaller 
Hubbard repulsion,  because both exhibit long-range antiferromagnetic ordering 
in quasi two-dimensional layers \cite{ircu1,ircu2}. 
The realization of an effective layered half-filled Hubbard model in Sr$_{2}$IrO$_{4}$ is a result of a spin-orbit 
coupling $\lambda$ close to 0.5~eV that splits the $t_{2g}$ states 
into $j_{\textrm{eff}}=1/2$ and $j_{\textrm{eff}}=3/2$ sectors with a gap approximately equal to $3\lambda/2$. At $n=5$ this 
leads to a half-filled $j_{\textrm{eff}}=1/2$ band and concomitant Mott/Slater insulator behavior. Besides the
iridates, other materials have also similarly interesting properties \cite{other1,other2,other3}.  
Even in the context of iron superconductors the importance of spin-orbit coupling has been remarked \cite{fe1,fe2,fe3,fe4}.

Another interesting scenario which has been recently theoretically investigated  led 
to the prediction of unusual magnetism in the $n=4$ case \cite{Khaliullin,meetei,Svoboda01,Sato02,exc1,exc2,Kaushal01}. 
At this electronic density, {\it spin-orbit excitons} (for details see Sec.~\ref{EXC_COND})
were found to condense at momentum $q=\pi$, both in the intermediate and strong coupling 
limits, and also display antiferromagnetic staggered magnetic order. Experimentally, for double perovskite materials 
such as Sr$_{2}$YIrO$_{6}$ and Ba$_{2}$YIrO$_{6}$, with Ir$^{5+}$ ions and a $5d^{4}$ configuration, the
presence of the exciton condensate, as discussed time ago in semiconductors \cite{rice1960}, has been debated \cite{Cao01,Corredor01,Terizc01,TDey01,Nag01}. 
Recent RIXS (resonant inelastic x-ray scattering)
experiments on Sr$_{2}$YIrO$_{6}$ and Ba$_{2}$YIrO$_{6}$ have unveiled $J_{\textrm{eff}}=1$ 
and $J_{\textrm{eff}}=2$ excitations  with weak dispersion at energies appproximately $0.37$~eV and $0.7$~eV \cite{Kusch01}, 
respectively, which suggests that the bandwidth of excitonic excitations is not sufficiently large when compared
with  $\lambda$ to realize the predicted spin-orbit exciton condensate. 
It should be noted that these $J_{\textrm{eff}}=1$ 
and $J_{\textrm{eff}}=2$ excitations can be understood in terms of more conventional excitonic (electron-hole pair) 
states \cite{rice1960} between $j_{\textrm{eff}}=3/2$ and $j_{\textrm{eff}}=1/2$ sectors. Because 
in these excitations electrons jump from $j_{\textrm{eff}}=3/2$ to $j_{\textrm{eff}}=1/2$ states, 
the addition of angular momentum suggests that this will lead 
to $J_{\textrm{eff}} \in \{\frac{3}{2}-\frac{1}{2}, \frac{3}{2}+\frac{1}{2}\}$ i.e.  $J_{\textrm{eff}}=1$ or $2$ excitations.
In the layered Sr$_{2}$IrO$_{4}$ compound, these spin-orbit excitons are also present as stable excited states as shown by recent  RIXS
and optical conductivity measurements~\cite{Souri01,Kim02,Kim01} 
[note that the notation $J_{\textrm{eff}}$ is used for the total effective angular momentum of the system (or an atom), 
while $j_{\textrm{eff}}$ refers for the effective angular momentum of single particle states. In the rest of the paper, we follow the same convention].

In addition to the above mentioned progress, it should be remarked that there are several real
quasi-one dimensional materials with robust spin-orbit coupling strength that have been studied in the literature. 
The doped variants of the materials reported below may directly realize the physics 
discussed in this publication, because our calculations are based on numerically exact solutions
of one-dimensional multiorbital models. For example, recently 1D stripes of Sr$_{2}$IrO$_{4}$ \cite{Gruenewald01} 
were grown epitaxially and RIXS spectra have shown the presence of spin-orbit excitons at energies nearly 0.6~eV.  
Other examples of one-dimensional $j_{\textrm{eff}}=1/2$ antiferromagnets includes CaIrO$_{3}$~\cite{Bogdanov01,Ohgushi01,Sala01,SWKim01}  
and Ca$_{4}$IrO$_{6}$~\cite{Cao02,Calder01}. BaIrO$_{3}$ also belongs to the $5d^{5}$ class but have shown an
unexpected charge-density wave~\cite{KMaiti01,OBKorneta01}.  There are also mixed $3d$-$5d$ one-dimensional insulators, such as 
Ba$_{5}$CuIr$_{3}$O \cite{MYe01} and Sr$_{3}$CuIrO$_{6}$ \cite{XLiu01,WGYin01}. La$_{3}$OsO$_{7}$, which lies 
in the category of $5d^{3}$, is also a quasi-one-dimensional material with antiferromagnetic ordering 
and $T_{\rm N}=45$~K~\cite{RMorrow01}. There are also examples of quasi-one-dimensional materials with fractional 
valence states of the Ir and Rh ions, such as Ba$_{5}$AlIr$_{2}$O$_{11}$~\cite{Terzic01,Streltsov01,YWang01}, Ca$_{5}$Ir$_{3}$O$_{12}$~\cite{XChen01,KMatsuhira01}, 
and Sr$_{3}$Rh$_{4}$O$_{12}$~\cite{Cao02}. BaRu$_{6}$O$_{12}$ and KRu$_{4}$O$_{8}$ are 
examples of quasi-one-dimensional ruthenates ~\cite{ZQMao01,Kobayashi} that have also attracted considerable attention.
The combination of the existence of real quasi-one-dimensional $4d$ and $5d$
materials  and our model studies employing numerically very accurate techniques provides a unique opportunity to 
explore and understand the phases which can emerge from the interplay of spin-orbit coupling, Coulomb electronic
repulsion, and kinetic energy.

To obtain our results we use the numerically accurate density matrix renormalization group (DMRG) technique \cite{dmrg1,dmrg2} to solve 
the degenerate three-orbital Hubbard model in one-dimension. Up to now studies of the phases 
emerging in fractionally-filled  three-orbital Hubbard models with spin-orbit coupling are relatively few, particularly
as compared to the thoroughly investigated integer fillings, such as $n=5$ and $n=4$. 
To develop a conceptual understanding, here we used doping $n=3.5$, i.e. 3.5 electrons per site in average, using 
a model with degenerate bands. Via DMRG calculations here we report
the phase diagram varying $\lambda$ and $U$. To our best knowledge, theoretical studies
at this electronic doping have not been presented before.

The organization of this paper is as follows. In Sec. II, the model used for our study 
is defined and the details of the computational method are explained. The main results are 
presented in Sec.~III, including the phase diagram varying $U$ and $\lambda$. In particular, 
firstly we present the evidence for the novel block excitonic phase that we unveiled, and then we address 
the different magnetic phases present in the complete phase diagram, followed by a description of the 
density of states (DOS). In Sec.~IV, we discuss our main results and present our conclusions.

\section{Model and Method}

In this manuscript, we use
the three-orbital Hubbard model in the presence of spin-orbit coupling. The Hamiltonian contains a tight-binding term, 
an on-site Hubbard-Hund interaction, and a spin-orbit coupling: $H = H_{K} + H_{\mathrm{int}} + H_{SOC}$. The tight-binding portion is 
\begin{equation}
H_{K} = -\sum_{{i},\sigma,\gamma,\gamma^{\prime}}t_{\gamma\gamma^{\prime}}
(c_{{i}\sigma\gamma}^{\dagger}c^\pdag_{{i}+1\sigma\gamma^{\prime}}+\mathrm{h.c.})
+\sum_{{i},\sigma,\gamma}\Delta_{\gamma}n_{{i}\sigma\gamma}.
\end{equation}
To gain conceptual understanding, we have focused on the simplest case of degenerate $t_{2g}$ states, 
hence we fixed $t_{\gamma\gamma^{\prime}}= t\delta_{\gamma\gamma^{\prime}}$, where $t=0.5$, and $\Delta_{\gamma}=0$. 
This leads to a total bandwidth ($W$) = 2.0 eV in the non-interacting limit. The on-site Hubbard-Hund interaction is
\begin{multline}\label{INT_term}
H_{\mathrm{int}} = U\sum_{{i},\gamma} n_{{i}\uparrow\gamma}
n_{{i}\downarrow\gamma} 
+\left(U'-J_{H}/2\right)\sum_{{i},\gamma<\gamma'} n_{{i}\gamma}
n_{{i}\gamma'} 
\\
  -2J_{H}\sum_{{i},\gamma<\gamma'} \mathbf{S}_{{i}\gamma} \cdot 
  \mathbf{S}_{{i}\gamma'} 
+J_{H}\sum_{{i},\gamma<\gamma'} \left( P^{\dagger}_{{i}\gamma} 
P_{{i}\gamma'} + \mathrm{h.c.} \right) . 
\end{multline}
In the above expression $n_{{i}\gamma}$ is the electronic 
density at orbital $\gamma$ and lattice site ${i}$, while the operator $\mathbf{S}_{{i}\gamma}$=${{1}\over{2}}\sum_{\alpha,\beta} 
c_{{i}\alpha\gamma}^{\dagger} \sigma_{\alpha\beta} c^\pdag_{{i}\beta\gamma}$ is 
the total spin. The first two terms describe the intra- and inter-orbital electronic repulsion, respectively. 
The third term contains the Hund coupling that favors the ferromagnetic alignment of the spins at different orbitals; the fourth term is the 
pair hopping with $P_{{i}\gamma}=c_{{i}\downarrow\gamma}c_{{i}\uparrow\gamma}$ 
as the pair operator. We assume the standard relation $U^{\prime}=U-2J_{H}$ based on rotational invariance, and we fixed $J_{H}=U/4$ as in \cite{Kaushal01}. 
Hence, only $U$ and $\lambda$ are free parameters in our study.

The SOC term is 
\begin{equation}\label{SO_term}
H_{\mathrm{SOC}}=\lambda\sum_{{i},\gamma,\gamma^{'},\sigma,\sigma^{'}}
{{\langle \gamma|{\bold{L}_{i}}|\gamma^{'}\rangle}\cdot{\langle\sigma|{\bold{S}_{i}}|\sigma^{'}\rangle}}
c_{i\sigma\gamma}^{\dagger}c_{i\sigma^{'}\gamma^{'}} \hspace{0.1cm},
\end{equation} 
where $\lambda$ is the SOC coupling strength.  

In the non-interacting limit, both the SOC and tight-binding terms can be diagonalized simultaneously to obtain the following Hamiltonian:
\begin{multline}\label{jj_coupling}
H_{K} + H_{SOC} =\sum_{k,m} (2t\cos(k)-\frac{\lambda}{2}){a_{k,\frac{3}{2},m}^{\dagger}}a_{k,\frac{3}{2},m} + 
\\
\sum_{k,m} (2t\cos(k)+\lambda){a_{k,\frac{1}{2},m}^{\dagger}}a_{k,\frac{1}{2},m}. 
\end{multline}
Above we used $a_{k,j_{\textrm{eff}},m}^{\dagger}=1/\sqrt{L}\sum_{l}e^{-\iota lk}a_{l,j_{\textrm{eff}},m}^{\dagger}$, 
where $a_{l,j_{\textrm{eff}},m}^{\dagger}$ is the creation operator for an electron with total effective angular momentum $j_{\textrm{eff}}$ and $z$-projection $m$. 
The transformation between the $t_{2g}$ orbitals and $j_{\textrm{eff}}$ basis is the following (real-space site index $l$ is dropped):
\begin{equation}\label{transformation}
\begin{bmatrix}a_{\frac{3}{2},\frac{3s}{2}}\\a_{\frac{3}{2},-\frac{s}{2}}\\a_{\frac{1}{2},-\frac{s}{2}}\end{bmatrix}
= \begin{bmatrix}\frac{is}{\sqrt{2}}&\frac{1}{\sqrt{2}}&0\\\frac{s}{\sqrt{6}}&\frac{i}{\sqrt{6}}&\frac{2}{\sqrt{6}}\\
\frac{-s}{\sqrt{3}}&\frac{-i}{\sqrt{3}}&\frac{1}{\sqrt{3}}
\end{bmatrix}\begin{bmatrix}c_{\sigma yz}\\c_{\sigma xz}\\c_{\bar{\sigma} xy}\end{bmatrix} ,
\end{equation}
where $s$ is $1(-1)$ when $\sigma$ is $\uparrow(\downarrow)$ and $\bar{\sigma}=-\sigma$ 
[note that from now onwards to avoid complicated notations, when $j_{\textrm{eff}}$ should be used as subindex, 
sometimes this quantum number will be simply denoted by $j$]. 
Equation~(\ref{jj_coupling}) is useful to understand the non-interacting limit of the model. As $\lambda$ is increased, the 
$j_{\textrm{eff}}=3/2$ and $j_{\textrm{eff}}=1/2$ bands split. For the doping $n=3.5$ addressed in this study, in the large $\lambda$ limit 
all electrons will be located in the $j_{\textrm{eff}}=3/2$ band, making the $j_{\textrm{eff}}=3/2$ band fractionally filled and the $j_{\textrm{eff}}=1/2$ band empty. 
This region is called $j_{\textrm{eff}}=3/2$ metal, as discussed in Sec.~\ref{LDDS}.

Because our primary interest is to understand the subtle phases emerging from the competition of the Coulomb interaction, spin-orbit coupling, and kinetic energy, 
we used the DMRG technique which is numerically exact in one dimension. DMRG can treat the above three terms 
in the Hamiltonian on equal footing. We solved the  above described model for various system lengths, such as $L=8,16,24,32,$ and $48$, fixing 
the average local density to $n=3.5$. To reduce the cost of the simulations, we have targetted subspaces of the total $J_{z}^{\textrm{eff}}=\sum_{i}(J_{z}^{\textrm{eff}})_{i}$,
which is possible because $[H,J^{\textrm{eff}}_{z}]=0$ for the chosen tight-binding parameters (for details see \cite{Kaushal01}). 
For the DMRG process, we used up to 1000 states and the corrected single-site DMRG algorithm~\cite{whiteSc} with correction $a=0.001-0.008$.
We performed 35 to 40 sweeps to gain proper convergence 
to the ground state properties. After convergence, we calculated the spin structure factor $S(q)$, local occupations $\langle n_{jm}\rangle$,  
local moments $\mathbf{S}^{2}_{i}$, $\mathbf{L}^{2}_{i}$, and $(\mathbf{J}^{\textrm{eff}}_{i})^{2}$, and also the exciton pair-pair 
correlation $\langle\Delta^{\dagger \tilde{j}\tilde{m}^{\prime}}_{j\tilde{m}}(i)\Delta^{\tilde{j}m^{\prime}}_{jm}(i^{\prime})\rangle$ in order 
to construct the phase diagram. Moreover, we also used the DMRG correction vector method \cite{kuhner99} with $L=16$,
as well as the Lanczos algorithm~\cite{Dagotto01} with $L=4$, to calculate the single-particle DOS. 

\section{Results}
\begin{figure}[!t]
\hspace*{-1.0cm}
\vspace*{0cm}
\begin{overpic}[width=1.2\columnwidth]{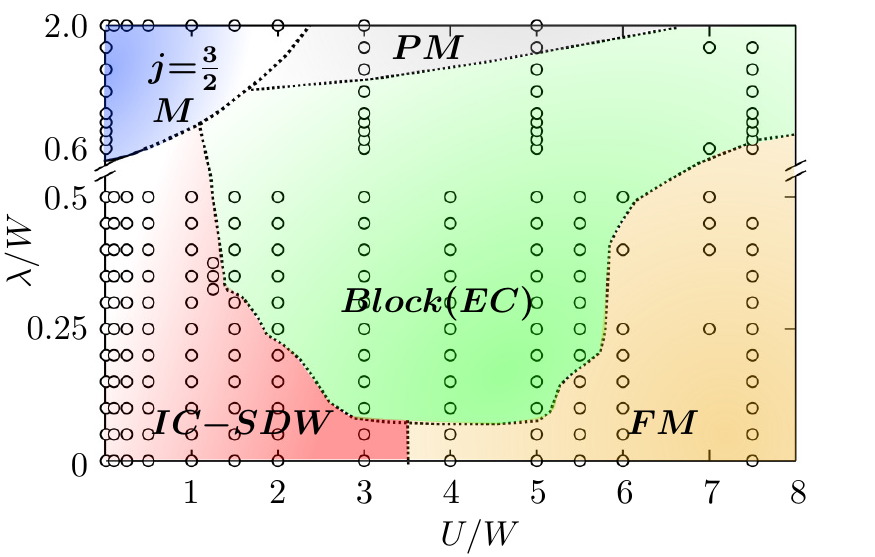}
\end{overpic}
\caption{Shown is the main result of our publication, namely the $\lambda$ - $U$ phase diagram for $n=3.5$, where $W=2.0$~eV.  
IC-SDW, EC, FM, and PM stands for  Incommensurate Spin Density Wave, 
Excitonic Condensate, Ferromagnetic, and Paramagnetic phases, respectively. The high density of points was achieved by using a
system size $L=16$ for the DMRG calculations, but several points in the phase diagram were obtained with $L=32$ chains, as described for special cases below.}
\label{fig1}
\end{figure}
In Fig.~\ref{fig1}, we show the phase diagram that we obtained varying $U$ and $\lambda$ in units of the 
non-interacting bandwidth $W$, at a fixed average local electronic density $n=3.5$. 
The main result is the presence of a ``block excitonic condensate'', accompanied with block magnetic ordering 
(in a $\uparrow\uparrow\downarrow\downarrow$ pattern). We  will discuss this novel phase, and other phases, in the following subsections. 
Also note that our study is in one dimension. For this reason when we express that in a range of $U$ and $\lambda$ 
we are located at a particular phase with particular characteristics, this always has to be interpreted in the sense of 
dominant power-law decaying correlations, as opposed to true long-range order. 
Hence, the excitonic condensation we focus on is actually a quasi-excitonic condensation in a one-dimensional system.

\subsection{Condensation of spin-orbit excitons}\label{EXC_COND}

We now proceed to show and discuss the evidence of excitonic condensation in our phase diagram. We define the creation 
operator for an exciton at site $i$ as $\Delta^{\dagger \tilde{j}m^{\prime}}_{jm}(i)=a^{\dagger}_{ijm}a_{i\tilde{j}m^{\prime}}$, 
where $j=1/2$ and $\tilde{j}=3/2$ are fixed. The exciton created by the above operator consists of a hole located 
at a $\tilde{j}=3/2$ state with projection $m^{\prime}$ 
and an electron with $j=1/2$ with projection $m$. These excitons are called ``spin-orbit 
excitons'' because the electron-hole pair is present in a spin-orbit entangled state. 
A similar excitonic operator was used before in \cite{Kaushal01,Sato02} to investigate the condensation 
of spin-orbit excitons for the $n=4$ case. We would like to mention that the present work is the first study 
where the condensation of these excitons is shown to be stable for $n=3.5$.
\begin{figure}[!t]
\hspace*{-1.5cm}
\vspace*{0cm}
\begin{overpic}[width=1.1\columnwidth]{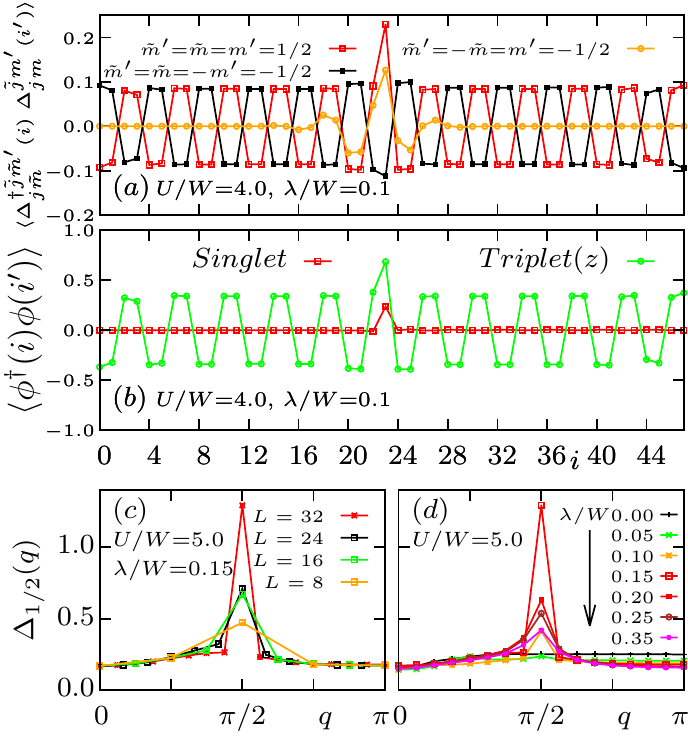}
\end{overpic}
\caption{Panels (a,b) show the exciton-exciton correlation in real space, at $U/W = 4.0$ and $\lambda/W = 0.1$ using a $L=48$ system size. 
For panels (a,b), $m=1/2$ and $i^{\prime}$=$23$ (in the middle of a chain with open boundary conditions) are fixed. 
Panel (c) shows the momentum distribution function for excitons at $U/W = 5.0$ and $\lambda/W = 0.15$. 
Similar momentum distribution functions of excitons for various $\lambda$'s are shown in panel (d). A system size $L=32$ is used for panel (d).}
\label{fig2}
\end{figure}

 As our calculations are performed for a large but finite system, 
we expect $\langle\Delta^{\dagger \tilde{j}m^{\prime}}_{jm}(i)\rangle=0$. 
Non-zero values of the above observable would imply that $U(1)$ symmetries,
corresponding to conservation of number of particles in the $j=1/2$ and $j=3/2$ states separately, 
are spontaneously broken. Hence to investigate excitonic condensation 
we measure the real-space correlation between the excitons 
i.e. $\langle\Delta^{\dagger \tilde{j}\tilde{m}^{\prime}}_{j\tilde{m}}(i)\Delta^{\tilde{j}m^{\prime}}_{jm}(i^{\prime})\rangle$. 
Here we would like to mention that earlier similar type of analysis was performed for one-dimensional systems 
to investigate quasi-excitonic condensations but in simpler models such as the extended Falicov-Kimball model \cite{Ejima01}.

For the spin-orbit excitons, as  the quantum number $m$ can take two values ($m=\pm1/2$), this gives rise to two channels 
for excitonic condensation, namely the singlet and triplet channels \cite{rice1960,Kunes01}. 
We define the exciton creation operators in both channels in the following manner:
\begin{eqnarray}\label{stChannel}
\phi^{s}(i) = \sum_{m} \Delta_{jm}^{\tilde{j}m \dagger}(i),
\\
\label{stChannel2}
\boldsymbol{\phi}^{t}(i) = \sum_{m m^{\prime}} \Delta_{j{m^{\prime}}}^{\tilde{j}m \dagger}(i) \boldsymbol{\tau}
_{m m^{\prime}},
\end{eqnarray}
\\
where $\boldsymbol{\tau}$ are the Pauli matrices.

In Fig.~\ref{fig2}(a), we show the real-space correlations between the exciton pairs (with respect to the central site defined 
as $i^{\prime}=23$) for a system size $L=48$, at $U/W=4.0$ and $\lambda/W=0.1$. In Fig.~\ref{fig2}(a), we fix $m=1/2$. 
Note the robust block ordering [$++--$] that we found in the pair-pair correlation between excitons
(see ``${\tilde{m}^{\prime}}=\tilde{m}=m^{\prime}=1/2$'', red line), consisting 
of holes and electrons with the same projection $j_{z}^{\textrm{eff}}=1/2$.  We also noticed the presence of very long distance correlations 
between excitons of opposite projections (see ``${\tilde{m}^{\prime}}=\tilde{m}=-m^{\prime}=-1/2$'' in Fig.~\ref{fig2}(a), black line), 
i.e. one exciton consists of a hole and electron pair with projection $j_{z}^{\textrm{eff}}=1/2$ and the other exciton is made up of a
hole and electron with projection $j_{z}^{\textrm{eff}}=-1/2$. Note that the above discussed excitonic correlations 
will contribute to both the singlet and the $z$ component of the triplet channels.
Similarly, we can create excitons consisting of electron ($j_{z}^{\textrm{eff}}=1/2$) and hole ($j_{z}^{\textrm{eff}}=-1/2$) with different projections:
the correlations between these excitons are shown in Fig.~\ref{fig2}(a) 
(see ``${\tilde{m}^{\prime}}=-\tilde{m}=m^{\prime}=-1/2$'', orange line) and they display a rapid exponential decay.
These excitonic correlations contribute to the $x$ and $y$ components of the triplet channel. 

Using the above information and Eqs.(\ref{stChannel},\ref{stChannel2}), we calculate the real-space correlations 
for the excitons in the singlet and triplet channels ($z$ component). As shown in Fig.~\ref{fig2}(b), clearly the 
triplet channel is the dominant showing quasi long-range order (likely a very slow power-law decay in our finite one dimensional system). 
As discussed before, the $x$ and  $y$ components of the triplet channel have exponential decay.
This asymmetry between the $x,y$, and $z$ components is just a consequence of targeting the 
total $J_{z}^{\textrm{eff}}$ sector in our DMRG simulations. We suspect in higher dimensional systems, the long-range 
correlations between these excitons will be accompanied by the breaking of $U(1)$ symmetries leading to the 
formation of a nonzero order parameter $\langle\Delta^{\dagger \tilde{j}m}_{jm}(i)\rangle \ne 0$ (and $\langle \phi_{z}^{t}(i) \rangle \ne 0$). 
But for the one-dimensional case studied here, long range correlations (slow power-law decays) is used as evidence for  
excitonic condensation. For a simpler two-band models, early work \cite{rice1960} showed 
that condensation of excitons in the triplet channel leads to a spin-density wave and in the singlet channel leads, instead, 
to a charge-density wave. 

Surprisingly, we also observed block magnetic ordering in the excitonic condensate phase reported here,
which will be discussed  in detail in Sec.~\ref{MOMS}. The dominating correlations in the $z$ direction of the 
triplet channel implies that the relevant excitons are created by pairing electron and holes with the 
same $j_z^{\textrm{eff}}$, and from now on we will focus only on these excitons.
 
The momentum distribution function for excitons is
$\Delta_{m}(q)=\frac{1}{L}\sum_{i,i^{\prime}}\langle\Delta^{\dagger \tilde{j}m}_{jm}(i)\Delta^{\tilde{j}m}_{jm}(i^{\prime})\rangle e^{\iota q(i-i^{\prime})}$.
This quantity provides an indication of the number of excitons (with projection $m$) at momentum $q$. 
In Fig.~\ref{fig2}(d) we show $\Delta_{m}(q)$ for $U/W=5$ and various $\lambda$'s.  
For spin-orbit coupling strength $\lambda/W \lessapprox 0.07$, in the ferromagnetic region (to be discussed in more detail in 
the next subsection), the momentum distribution function is nearly flat. But at larger spin-orbit coupling, 
excitons condense at momentum $\pi/2$ and on further increasing $\lambda$, the number of excitons at $q=\pi/2$ again decreases, 
as shown in Fig.~\ref{fig2}(d). The spin-orbit coupling strength for this crossover from the 
ferromagnetic phase to the block-excitonic condensate depends on the strength of $U$. 
We noticed the interesting feature (see phase diagram in Fig.(\ref{fig1})) 
that for $U/W \gtrapprox 5$, a larger $U$ needs a larger $\lambda$ for the condensation to occur. 
We also found $\pi/2$ order in the excitonic correlations above the IC-SDW region, as shown in the phase diagram of Fig.~1, 
contrary to the strong coupling region, where increasing $U$ needs smaller a $\lambda$ 
for stabilizing the block-excitonic condensate.

We also show the finite-size scaling of the excitonic momentum distribution function in Fig.~\ref{fig2}(c), 
for system sizes $L = 8, 16, 24,$ and $32$. The nearly linear growth of $\Delta_{1/2}(q=\pi/2)$ with the system size ($L$) 
again suggests the presence of a robust excitonic condensation. 

\subsection{Magnetic ordering}\label{MOMS}

In this subsection, we will discuss and show the evidence for the different types of magnetic orderings 
found in the phase diagram. To investigate the various magnetic orderings, we calculate the spin-spin correlation 
$\langle\bold{S}_{i} \cdot \bold{S}_{j}\rangle$, and associated 
spin structure factor $S(q)=\frac{1}{L}\sum_{i,j}\langle\bold{S}_{i} \cdot \bold{S}_{j}\rangle e^{\iota(i-j)q}$. 
We also calculated the averaged local moments, $\langle \bold{S}^2 \rangle = \frac{1}{L}\sum_{i}\langle \bold{S}_{i}^2 \rangle$. 
Similarly we calculated $\langle \bold{L}^2 \rangle$, $\langle (\bold{J}^{\textrm{eff}})^2 \rangle$, and $\langle \bold{L} \cdot \bold{S} \rangle$. 
To evaluate the angle between the average local spin and average local orbital moment, i.e. $\phi_{LS}$, 
we used $\phi_{LS}=cos^{-1}(\frac{\langle \bold{L} \cdot \bold{S} \rangle}{\langle l \rangle \langle s \rangle})$; where $\langle l\rangle (\langle l \rangle +1)=\langle \bold{L}^2 \rangle$ , and $\langle s\rangle ( \langle s\rangle+1)=\langle \bold{S}^2 \rangle$. 

\begin{figure}[!t]
\hspace*{-0.85cm}
\vspace*{0cm}
\begin{overpic}[width=1.1\columnwidth]{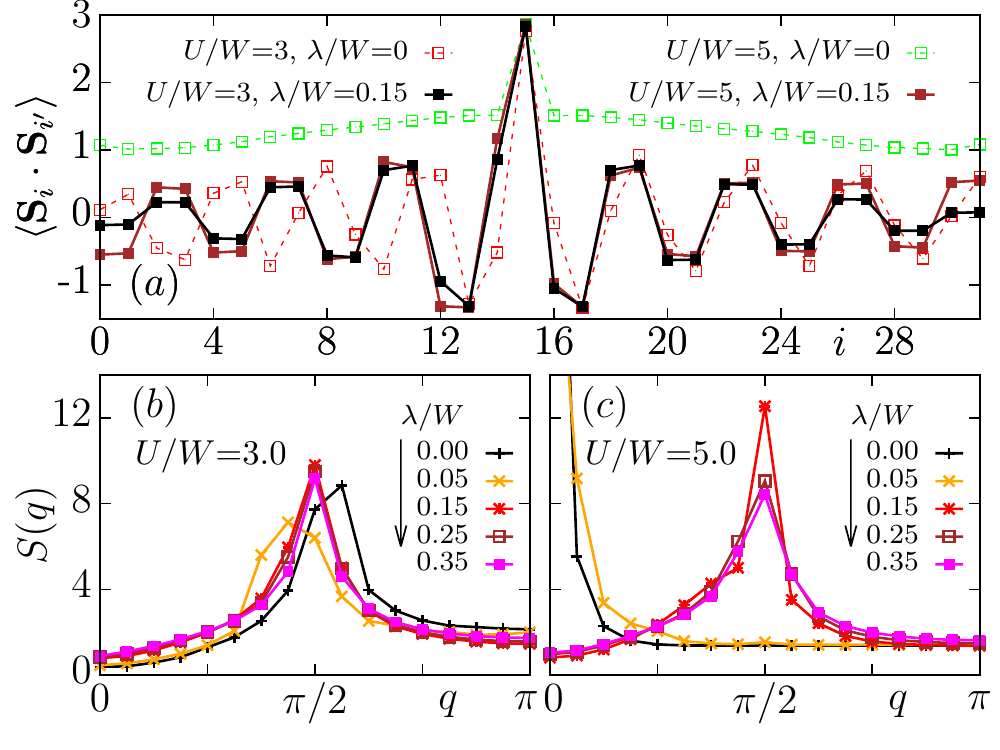}
\end{overpic}
\caption{In panel (a) the real-space spin-spin correlations are shown for $U/W=3$ and $U/W=5$, and for $\lambda/W=0$ 
and $\lambda/W=0.15$. In panels (b) and (c), the spin structure factor $S(q)$ is shown for  $U/W=3$ and $U/W=5$, 
respectively, and for various values of $\lambda/W$, as indicated.}
\label{fig3}
\end{figure}

 For $U/W \lessapprox 3.5$, we found that the IC-SDW region is smoothly connected to the non-interacting limit.
In this region (red region in Fig.~\ref{fig1}) the local moments gradually form up to the saturated values 
$\langle \bold{S}_{i}^{2}\rangle=2.875$ and $\langle \bold{L}_{i}^{2}\rangle=1$ as we increase $U/W$. Eventually this IC-SDW phase crossovers 
to the ferromagnetic (FM) phase. We noticed that in this IC-SDW region, the spin-ordering 
vector continuously changes depending on the values of  $U$ and $\lambda$, which we believe is the result of Fermi surface 
renormalization by the combined effect of $U$ and $\lambda$. In Fig.~\ref{fig3}(a), we show the real-space spin-spin correlation for $U/W=3$
 and $\lambda/W=0$ depicting the incommensurate order, for a system size $L=32$. 
By increasing $\lambda$, we found the block magnetic ordering phase above this IC-SDW region, 
but only as long as $U/W \gtrapprox 1.0$,  as shown in the phase diagram Fig.~\ref{fig1}. 
This block magnetic order survives up to $\lambda/W \approx 1$. In Fig.~\ref{fig3}(b), we display the spin structure
factor $S(q)$ for $U/W=3$ for various $\lambda/W$'s depicting the clear transition from IC-SDW to block magnetic order. 
We also show the real-space spin-spin correlation at $U/W=3$ and $\lambda/W=0.15$ in Fig.~\ref{fig3}(a), 
portraying the block magnetic order $(\uparrow\uparrow\downarrow\downarrow\uparrow\uparrow$).

As shown in Fig.~\ref{fig1}, for $\lambda/W=0$ the above mentioned IC-SDW phase is directly connected 
to the ferromagnetic (FM) region in the strong coupling limit; this crossover happens approximately at $U/W \approx 3.5$. 
In this FM phase, the local spin and orbital moments are fully saturated. The saturated values of the moments can be 
understood by considering the two-sites case in the large $U$ limit. As we are interested in density $n=3.5$, 
in the large $U$ limit the main contribution to the two-sites ground state will arise 
from the $d^{3}$-$d^{4}$ configuration. The $d^{3}$ site will have 
local $\langle \bold{S}^{2}\rangle=\frac{3}{2}(\frac{3}{2} + 1)=3.75$, while the $d^{4}$ site will have local $\langle \bold{S}^{2}\rangle=2$, 
leading  to an average $2.875$. Similarly the average local moment $\langle \bold{L}^{2}\rangle=1$ arises 
from $\langle \bold{L}^{2}\rangle=0$ (for $d^{3}$) and $\langle \bold{L}^{2}\rangle=2$ (for $d^{4}$).

\begin{figure}[!t]
\hspace*{-0.3cm}
\vspace*{0cm}
\begin{overpic}[width=1.1\columnwidth]{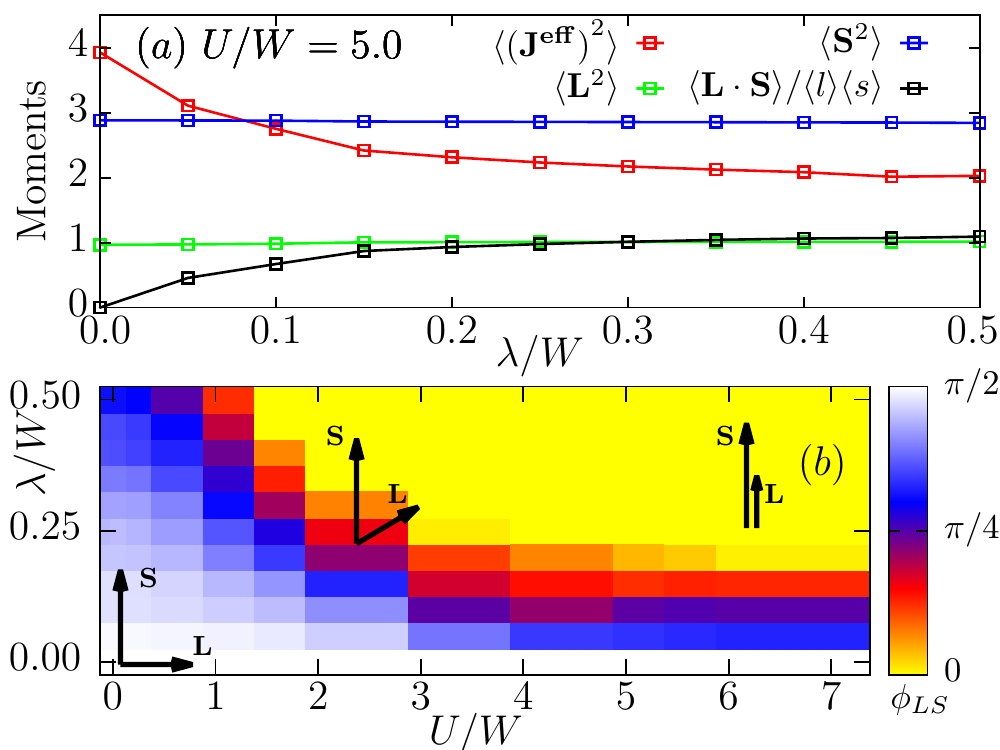}
\end{overpic}
\caption{(a) Average local moments and $\langle \bold{L} \cdot \bold{S}\rangle/\langle l \rangle\langle s\rangle$ for $U/W=5$, shown for various $\lambda/W$s. 
In panel (b), $\phi_{LS}$ is presented for several $\lambda/W$'s and $U/W$'s, where the color depicts the value of $\phi_{LS}$ as shown in the 
side panel scale. }
\label{fig4}
\end{figure}

As we increase $\lambda$ in this FM phase, the system eventually transitions to the 
block magnetic ordering. In Fig.~\ref{fig3}(a), we show the real-space spin-spin correlations for $U/W=5.0$ and $\lambda/W=0.0$, 
and for $U/W=5.0$ and $\lambda/W=0.15$, as evidence of the ferromagnetic and block magnetic orders, respectively. 
The spin structure factors for $U/W=5.0$ are shown in Fig.~\ref{fig3}(c) for various $\lambda's$, depicting 
the crossover from the ferromagnetic to the block magnetic ordered phases. 
As discussed in Sec.~\ref{EXC_COND}, we suspect this block magnetic ordered phase 
is related with the condensation of spin-orbit excitons at momentum $q=\pi/2$. Note
that for all the points shown in Fig.~\ref{fig1} in the block exciton condensate region (green region in the 
phase diagram of Fig.~\ref{fig1}), we found block magnetic ordering and condensation of excitons at $q=\pi/2$. 
We also noticed that, as we increase $\lambda$, gradually $\Delta_{1/2}(q=\pi/2)$ decreases and the system transitions smoothly into  
a paramagnetic (PM) phase.

Now we turn our focus towards the effect of spin-orbit coupling on the local moments. In Fig.~\ref{fig4}(a), 
we fixed $U/W=5$, and on  increasing $\lambda$ we observed that the local $\langle \bold{S}^{2} \rangle$ and $\langle \bold{L}^{2} \rangle$ 
remain nearly $2.875$ and $1.0$, respectively. But although their magnitudes are nearly constant 
there is a substantial change in the relative orientation 
of the spin and orbital moments i.e. they gradually modify their relative angle from $\pi/2$ to $0$. 
This rotation affects the local $\langle(\bold{J}^{\textrm{eff}})^{2}\rangle$, which decreases as the 
spin and orbital moments become parallel. In Fig.~\ref{fig4}, we show the average angle between the 
local spin and orbital moments. We notice that, as we increase $U$, smaller $\lambda$'s are enough 
to render $S$ and $L$ parallel to each other. This indicates that {\textit{Coulomb interactions enhance 
the effect of spin-orbit coupling}} and helps to entangle the spin and orbital moments. 
It is interesting to observe that the novel block excitonic phase we found 
lies in the region where  $L$ and $S$ are parallel to each other with $\langle {{\bf L} \cdot {\bf S}} \rangle \approx 
\langle l \rangle \langle s \rangle \approx 0.8$. We would like to mention that $\langle {{\bf L}\cdot {\bf S}}\rangle$ 
is directly related to the branching ratio calculated by XAS (X-ray Absorption Spectroscopy) experiments 
for the materials where SOC is robust \cite{LMarco01}.

\subsection{Local Densities and Density of States}\label{LDDS}

In the last two subsections we established the presence of a novel block excitonic condensate, 
accompanied with block magnetic order. Now we will discuss the spin-orbit basis-resolved average 
local occupations $\langle n_{jm}\rangle = \frac{1}{L}\sum_{i}\langle n_{i,jm}\rangle$ using a $L=16$ sites system. We also calculate 
the DOS ($\rho_{jm}(\omega-\mu)$) on a 4-site chain using Lanczos \cite{Dagotto01}, 
and on a $L=16$ system using the DMRG correction vector method \cite{kuhner99}, where $\mu$ is the chemical potential evaluated
via $(E(N+1) - E(N-1))/2$ for a system with $N$ electrons. In particular, we used the \textsc{DMRG++}
computer program \cite{alvarez09} and the Krylov formulation \cite{Nocera01} for the
DMRG correction vector method \cite{kuhner99}. Details on these calculations are provided in \cite{supplemental}.
\begin{figure}[!t]
\hspace*{-0.5cm}
\vspace*{0cm}
\begin{overpic}[width=1.0\columnwidth]{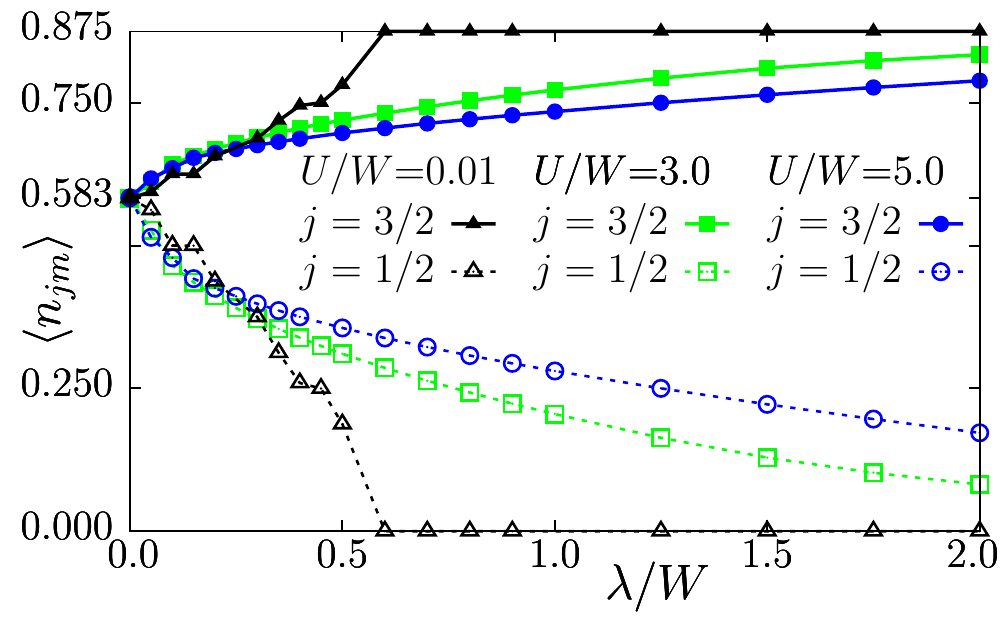}
\end{overpic}
\caption{Average local occupations $\langle n_{jm}\rangle$ shown for $U/W=0.01$,  $U/W=3.0$, and $U/W=5.0$, with increasing $\lambda/W$.}
\label{fig5}
\end{figure}

\begin{figure}[!t]
\hspace*{-1.0cm}
\vspace*{0cm}
\begin{overpic}[width=1.1\columnwidth]{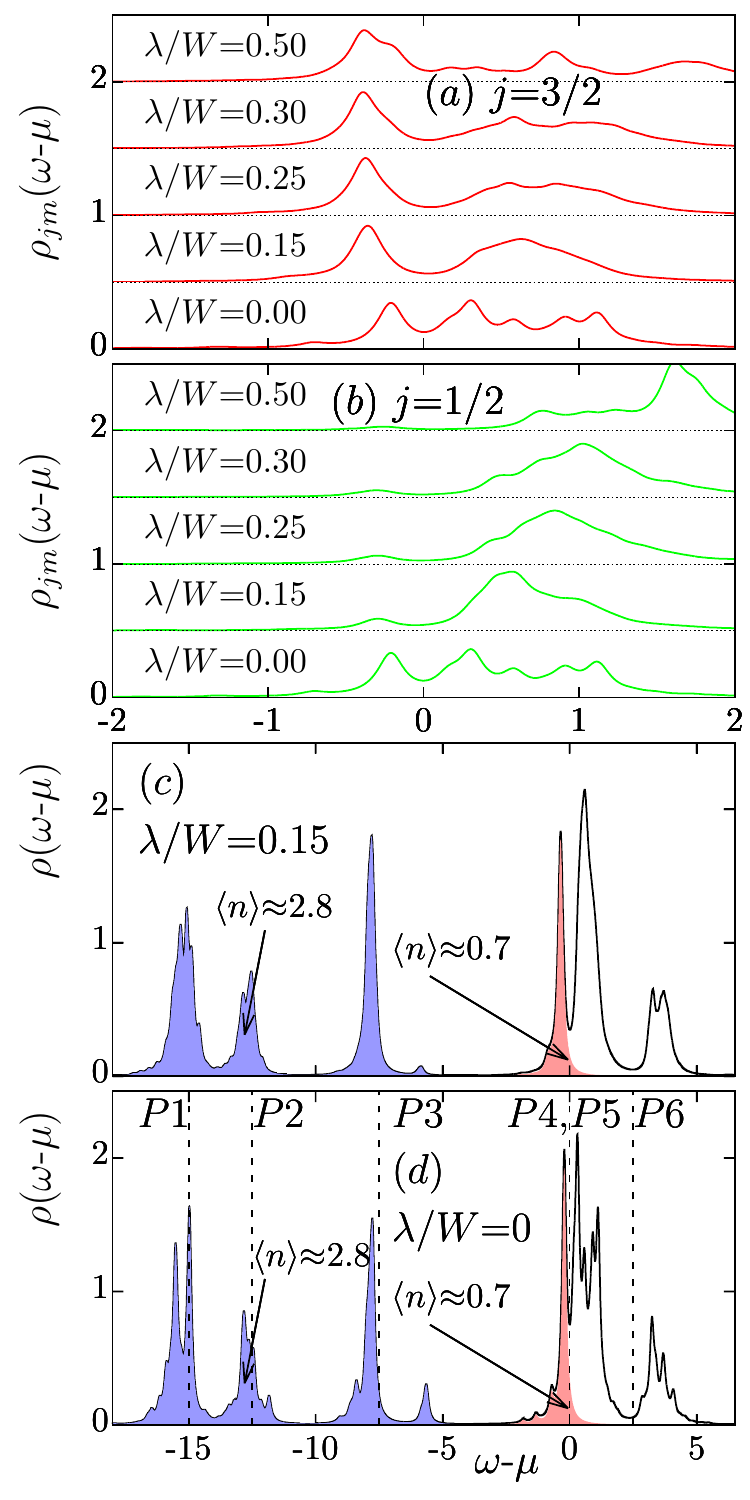}
\end{overpic}
\caption{Density of states, $\rho_{jm}(\omega-\mu)$, near $\mu$ shown for various values of $\lambda/W$, and at $j_{\textrm{eff}}=3/2$ 
and $j_{\textrm{eff}}=1/2$ in panels $(a)$ and $(b)$, respectively. In panels (c) and (d), the total $\rho(\omega - \mu)$ is shown 
for $\lambda/W=0.15$ and $\lambda/W=0$, respectively. All the above results are calculated for a $L=4$ (OBC) site cluster using Lanczos at
$U/W=5.0$. A broadening $\eta=0.1$ was used for all the results above.}
\label{fig6}
\end{figure}

In Fig. \ref{fig5}, we show the effect of spin-orbit coupling on the average local occupations $\langle n_{jm} \rangle$ for  three 
different $U$ values i.e. $U/W=0.01$, $U/W=3.0$, and $U/W=5.0$. Before explaining the results we would like to mention 
that the occupations in the $t_{2g}$ orbital basis for any $\lambda$ and $U$ are found to be same i.e. 
$\langle n_{\sigma \alpha}\rangle \approx 0.5833$, which is consequence of using degenerate orbitals in the kinetic energy term, 
and Coulomb interaction and spin-orbital coupling that do not break this symmetry in the $t_{2g}$ orbitals. The ``good'' basis 
for systems in the presence of spin-orbit coupling (in non-interating limit) is provided by the $j,m$ states, thus 
it is reasonable to discuss the occupation in terms of $\langle n_{jm} \rangle$. 
For small values of Coulomb interaction, such as $U/W=0.01$, we clearly reproduce the physics of the non-interacting limit. 
For $\lambda=0$ we found  $\langle n_{jm} \rangle \approx 0.5833$ and as we increase $\lambda$ the system 
transitions to a $j_{\textrm{eff}}=3/2$ metallic regime where the low-energy $j_{\textrm{eff}}=3/2$ band is fractionally filled with 
$\langle n_{3/2 m} \rangle=0.875$ electrons per site, and the higher energy band $j_{\textrm{eff}}=1/2$ is empty. 
At larger $U/W$ values, this $j_{\textrm{eff}}=3/2$ metallic phase is pushed towards larger $\lambda$. 
As also shown in Fig.~\ref{fig5}, that  for $U/W=3$ and $U/W=5$ we do not find a $j_{\textrm{eff}}=3/2$ metal for $\lambda/W$ as large as 2.0. 
This explains the curvature of the lower boundary of the $j_{\textrm{eff}}=3/2$ metal in the phase diagram Fig.~\ref{fig1}. 
To confirm these results, we also calculated the local occupations for different $U/W$'s at a fixed $\lambda/W=2.0$, 
and found that increasing $U$ gradually increases the filling in the  $j_{\textrm{eff}}=1/2$ state.

Now let us discuss the DOS calculated using the Lanczos method employing a four-site cluster 
with open boundary conditions. We checked that even using such a small four-site system, we obtain 
the same phases as in the phase diagram Fig.~\ref{fig1}. Firstly, let us discuss the total DOS
$\rho(\omega -\mu) = \sum_{jm}\rho_{jm}(\omega -\mu)$ for $\lambda=0$, as shown in Fig.~\ref{fig6}(d) for $U/W=5.0$ 
where we have a FM ground state. We noticed that away from the chemical potential we have four dominant peaks, 
named $P1$, $P2$, $P3$, and $P6$. These single-particle excitations can be understood in the strong coupling limit using a two-site cluster, 
as explained in \cite{supplemental}. The interesting feature is the presence of a metallic band near the chemical potential:
in the two-site limit this band consists only of two single-particle excitations $P4$ and $P5$. This metallic band in the
strong coupling limit \cite{supplemental} contains nearly $0.5$ itinerant electrons per site, which are moving in the ferromagnetic background of the
other electrons, and leads to a FM metal. For $U/W=5$, the occupied part of this metallic band is made of nearly $0.7$ electrons per site. 

Let us investigate the effect of $\lambda$ on the density of states. In Fig.~\ref{fig6}(c), we show 
$\rho(\omega - \mu)$ for $U/W=5$ and $\lambda/W=0.15$. At these couplings our Lanczos results show block 
magnetic order (with block excitonic order). We noticed that the positions of the peaks away from $\mu$ are 
not changed much, but the DOS at  $\mu$ decreases with a tendency to open a gap as we move into the block excitonic
phase. To explore this issue further, we calculated the $j,m$-resolved DOS $\rho_{jm}(\omega -\mu)$ 
near the chemical potential for different $\lambda/W$ values, as shown in Figs.~\ref{fig6}(a,b). 
For both $j_{\textrm{eff}}=3/2$ and $1/2$, note that with increasing $\lambda/W$ the DOS near $\mu$ decreases  
with the split in the metallic band. But for $j_{\textrm{eff}}=1/2$, the DOS below $\mu$ gradually decreases 
to $0$, because in the limit of very large $\lambda/W$ the $j_{\textrm{eff}}=1/2$ states will be empty. It should be also noted 
that for $\lambda/W=0.5$, we already see the emergence of states near the chemical potential for $j_{\textrm{eff}}=3/2$, 
with the system moving towards the paramagnetic phase. From the above exact analysis of the four-site system, we 
can conclude that there is a clear tendency towards the opening of a gap in the block phase, thus a clear tendency to form an insulator.

\begin{figure}[!t]
\hspace*{-1.0cm}
\vspace*{0cm}
\begin{overpic}[width=1.1\columnwidth]{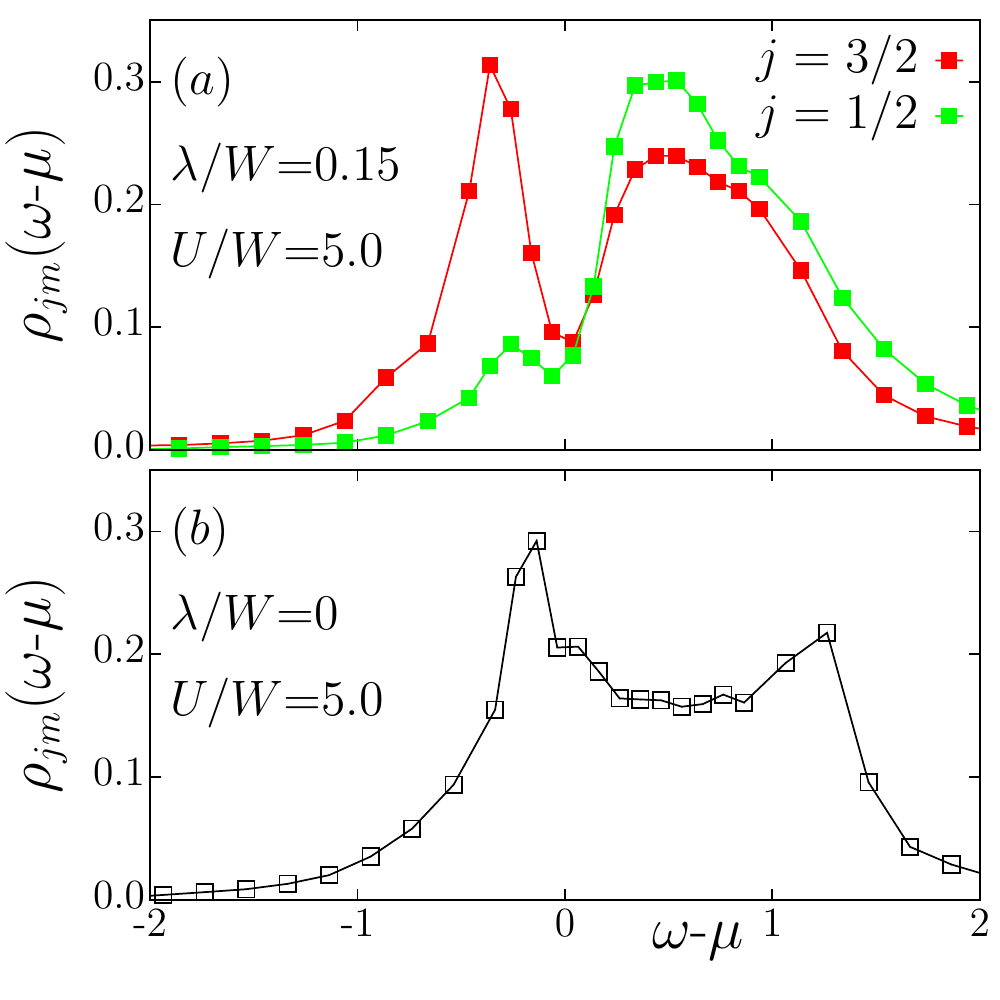}
\end{overpic}
\caption{Density of states, $\rho_{jm}(\omega-\mu)$, shown for a $L=16$ site system at 
$\lambda/W=0.15$ and $\lambda/W=0$ in panels (a) and (b), respectively. The DMRG vector correction method is used to calculate these results. 
$U/W=5.0$ and a broadening $\eta=0.1$ are employed. }
\label{fig7}
\end{figure}

To confirm that the above described results persists for larger systems, 
we also calculated $\rho_{jm}(\omega -\mu)$ near the chemical potential using the DMRG vector correction method for a $L=16$ site system. 
We again fixed $U/W=5$ and focused on $\lambda/W=0$ and $\lambda/W=0.15$, which shows ferromagnetic and block magnetic ordering, respectively,
as discussed in Sec.~\ref{MOMS}. We noted that the metallic band is clearly present in the ferromagnetic phase, 
see Fig.~\ref{fig7}(b). This suggests that indeed there is a fraction of electrons that develop a metallic band, 
having other localized electrons create a ferromagnetic background with spins $S \approx 3/2$ \cite{supplemental}. 
If now we increase $\lambda/W$ to 0.15, driving the system towards the block excitonic phase, both the $j_{\textrm{eff}}=3/2$ and $1/2$ sectors
show a tendency to open a gap and being insulating. These results further confirm the understanding deduced
from the small $L=4$ exact result. 

The novel block phase shown in this publication resembles 
the block magnetic order phase reported earlier in the context of three-orbital Hubbard models. 
In fact, block magnetic ordering, without the excitonic condensate component, has been found previously in models 
without spin-orbit coupling, in the context of the Orbital Selective Mott Phase (OSMP) \cite{Rincon01,Rincon02,Liu01,Li01,Herbrych01}.
In the latter, two orbitals are metallic with fractional filling and one orbital is insulating with half-filling. 
However, note that the block phase discussed in this publication is {\it not} accompanied by an OSMP phase, as indicated by the 
average local occupations. Instead the novel block phase discussed here is accompanied by the condensation of 
spin-orbit excitons at momentum $q=\pi/2$ for which a finite spin-orbit coupling is a necessary condition.

\section{Conclusions}
In this paper, we used an accurate numerical technique, DMRG, to construct the $\lambda$ vs $U$ phase diagram 
for the one-dimensional three-orbital Hubbard model at $n=3.5$. As our main result, we provide the 
first numerical evidence for the condensation of spin-orbit excitons in the fractionally filled three-orbital Hubbard model. 
Our calculations show that the spin-orbit excitons condense in the triplet channel and at momentum $\pi/2$, for all the points 
shown inside the green region of the phase diagram displayed in Fig.~\ref{fig1}. This quasi-condensation of excitons 
is accompanied by tendencies to open a gap at the chemical potential and also by block magnetic ordering. 
Interestingly, the block excitonic condensate unveiled here can be stabilized by introducing the spin-orbit coupling 
on both the IC-SDW and Ferromagnetic metallic phases. We also noticed that in this novel block excitonic condensate phase, 
local spin and orbital moments are highly entangled and nearly parallel to each other.

We believe the results reported in this publication -- which are unique given the considerable computational effort involved that requires robust
computational resources --  will encourage further theoretical and experimental investigations on 
fractionally-filled iridates~\cite{WJu01,Gunasekera01,Gunasekera02,Jayita,Panda} and also on other 
quasi-one dimensional materials with large spin-orbit coupling.  While our model calculations cannot establish which precise
material will realize the novel phase unveiled, we believe from now on the block condensate  
has to be considered among the candidate states when $n=3.5$ materials are studied.

\section{Acknowledgments}
N.K., A.N., A.M., and E.D. were supported by the U.S. Department of Energy (DOE), 
Office of Science, Basic Energy Sciences (BES), Materials Sciences and Engineering
Division. G.~A.~was partially supported by the Center for Nanophase
Materials Sciences, which is a DOE Office of Science User Facility,
and by the Scientific Discovery through the Advanced Computing (SciDAC)
program funded by U.S. DOE, Office of Science, Advanced Scientific
Computing Research and Basic Energy Sciences, Division of Materials Sciences and Engineering.



\newpage
\FloatBarrier


\end{document}


\begin{center}
{\bf \uppercase{Supplementary Information}} for\\
\vspace{10pt}
{\bf \large Novel Block Excitonic Condensate at n=3.5 in a \\
Spin-Orbit Coupled $t_{2g}$ Multiorbital Hubbard Model}\\
\vspace{10pt}
by N. Kaushal, A. Nocera, G. Alvarez, A. Moreo, and E. Dagotto
\date{\today}
\end{center}


\section{Two sites problem in strong coupling ($\lambda=0$)}

\begin{figure*} [b!]
\begin{tikzpicture}[]
\node (pic1) at (-2,0) {\includegraphics[width=0.62\linewidth, height=10cm]{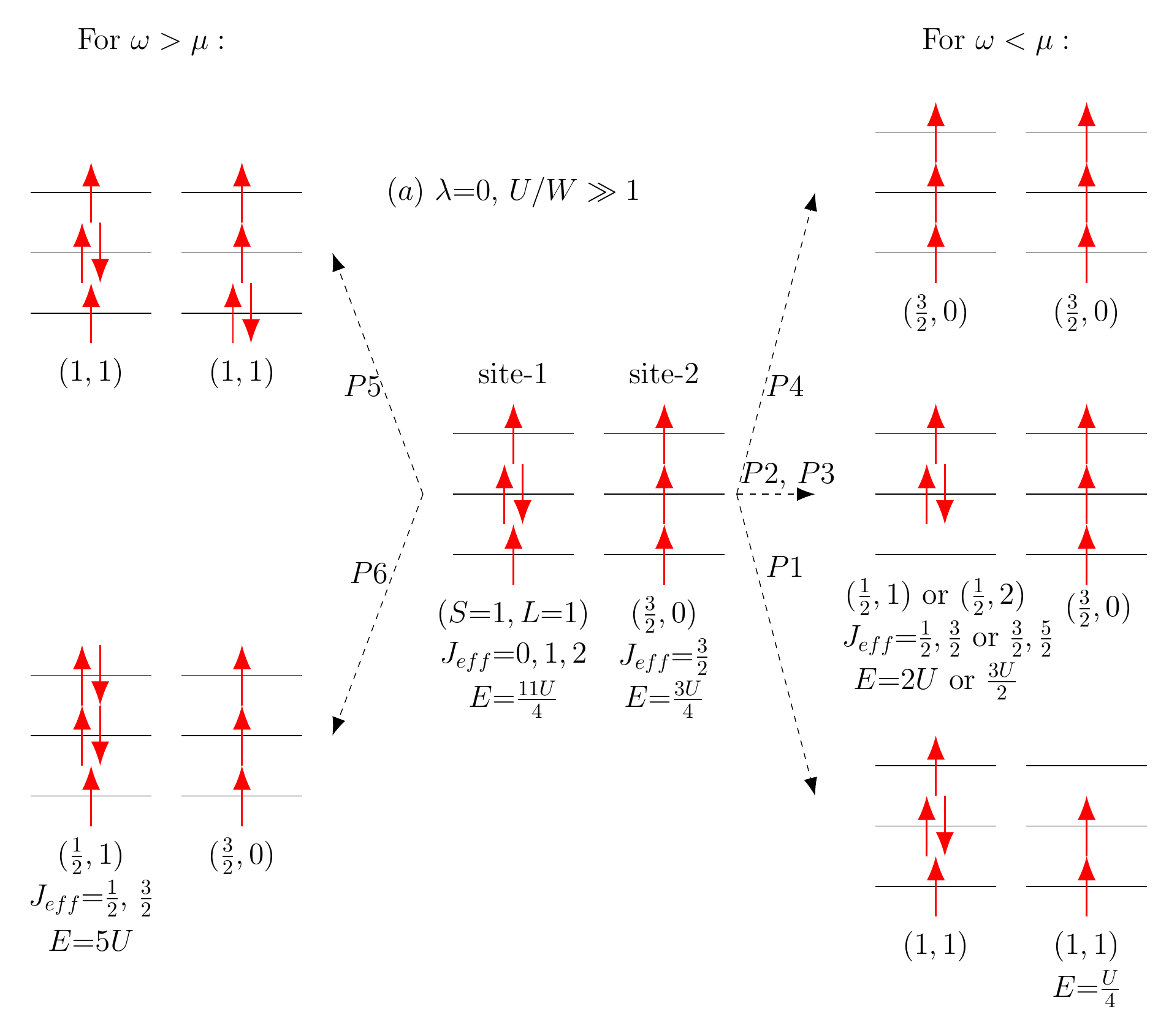}};
\node (pic2) at (7.7,2.1) {\includegraphics[width=0.45\linewidth, height=5.3cm]{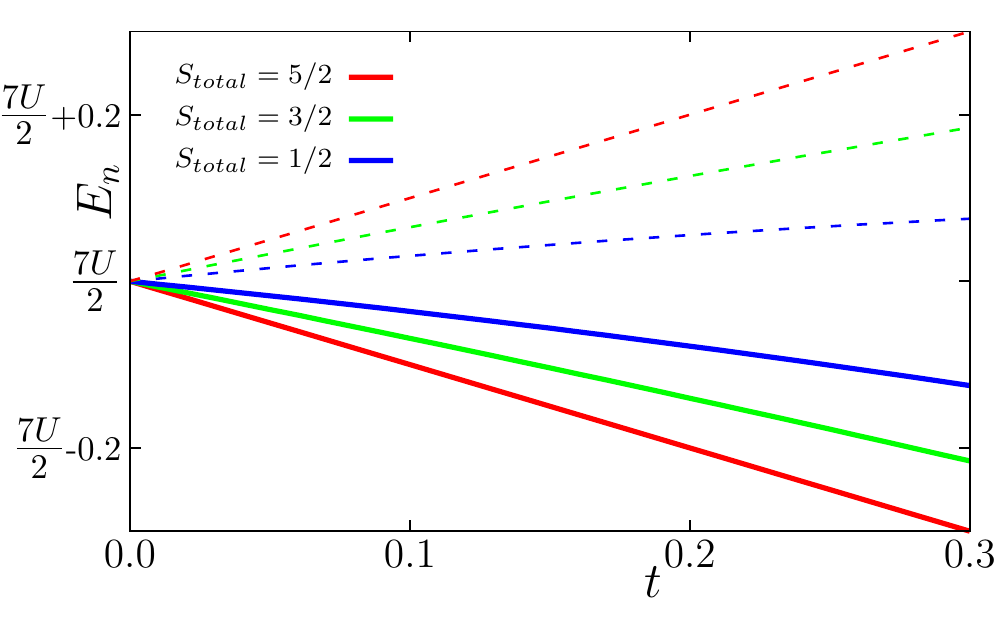}};
\node (pic2) at (7.7,-3.0) {\includegraphics[width=0.45\linewidth, height=5.3cm]{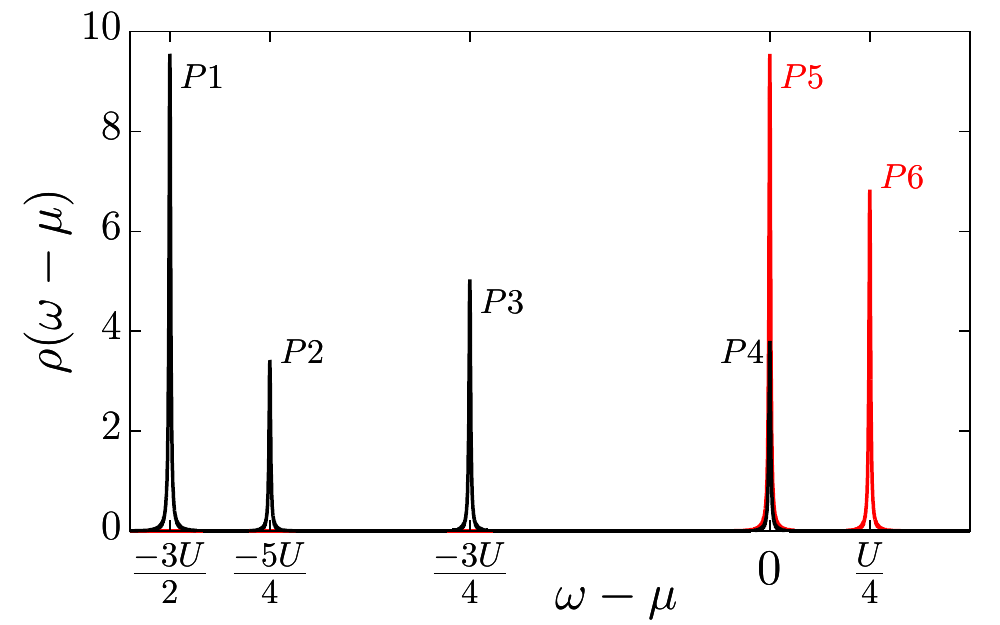}};
\node[text width=4cm] at (-0.5,-4) {(a)};
\node[text width=4cm] at (7.5,1) {(b)};
\node[text width=4cm] at (8,-2) {(c)};
\end{tikzpicture}
\centering
\caption{In panel (a), we provide a pictorial representation of the states leading to excitations in the single-particle density-of-states. 
Panel (b) shows the evolution of the low-energy states as the hopping is increased, for a 2-sites system in the strong coupling limit. 
In panel (c), $\rho(\omega)$ is shown for a 2-sites system in the strong coupling limit. A broadening $\eta=0.05$ is used. }
\end{figure*}\label{fig1} 


In this section, we discuss the 2-sites three-orbital Hubbard model in the strong coupling limit ($U/t>>1$) for $n=3.5$ i.e a total of 7 electrons. 
Before discussing further details, below we show the Coulomb interaction term of the Hamiltonian in the rotationally invariant form 
(we have used $U^{'}=U-2J_{H}$) which we use to calculate the energy of each atom given the occupation, spin moment ($S$), and orbital moment ($L$):
\begin{equation}\label{eqn1}
H_{int}=(U-2J_{H})\frac{N(N-1)}{2} + J_{H}(\frac{5}{2}N-2S^{2}-\frac{L^2}{2})
\end{equation}

If we work at $t=0$, atomic limit, the ground state (GS) of the 2-sites system consists only of $d^{3}$ and $d^{4}$ sites 
which have $(S,L)$=$(3/2,0)$ and $(1,1)$, respectively. Using Eq.(\ref{eqn1}), the total energy $E=E_{1}+E_{2}$  can be calculated to be $7U/2$. 
The total spin moment $S_{total}=S_{1} + S_{2}$ of the GS can be $\{1/2, 3/2, 5/2\}$ and the total orbital moment $L_{total}$ is 1. 

Now to understand the effect of a nonzero kinetic energy, we perform exact diagonalization of the 2-site clusters with finite hopping. 
As soon as we turn-on the hopping, we noticed the degeneracy in $S_{total}$ breaks, as shown in Fig.~\ref{fig1}(b). 
Interestingly all $S_{total}$ manifolds splits in their high-energy anti-bonding (dashed lines) and low-energy bonding (solid lines) states. 
The GS for finite hopping has a saturated total spin moment i.e. $S_{total}=5/2$, thus the GS has ferromagnetic ordering. 
The GS state has a degeneracy $6 \times 3=18$ fold, where ``$6$'' and ``$3$"  arise from $SU(2)$-symmetries in the spin sector 
of the $S=5/2$ and $L=1$ space. We also noticed that for $U/t>>1$, for all $S_{total}$ manifolds, the energies for the bonding(-) and anti-bonding(+) 
states change linearly with $t$ as $\frac{7U}{2} \pm \tilde{t}$, where the $\tilde{t}$ for $S_{total}=5/2$, $3/2$, $1/2$ manifolds 
are $t$, $\frac{2}{3}t$, and $\frac{1}{3}t$, respectively.

The above discussed bonding and anti-bonding states can be interpreted in terms of momentum $q=0$ and $q=\pi$ states, respectively, 
of the incipient bands which fully develops for the larger system and leads to band formation near the chemical potential. 
The presence of bands near the chemical potential is indeed confirmed by the density-of-state calculations performed on 4 and 16 sites systems 
using Lanczos and DMRG techniques, as shown in the main paper.  We also noticed that the density-of-states of the 2 sites system can explain 
most of the features present in $\rho(\omega)$ of 4-site clusters in strong coupling limit. In Fig.~\ref{fig1}(c) we show the $\rho(\omega-\mu)$ 
for the 2-site cluster in the strong coupling limit, and in Fig.~\ref{fig1}(a) we show all the ``N+1'' and ``N-1'' particle states leading 
to excitations above and below the chemical potential, respectively. For 4-site clusters we observed that these peaks are present nearly 
at the same values $\omega's$ in terms of $U$, as shown in the main paper.

\section{Single particle density-of-states}

As discussed in the main paper, we calculated the density-of-states $\rho_{jm}(\omega)$ using the 
DMRG correction-vector target method. The electron ($\omega<\mu$) and hole ($\omega>\mu$) components of $\rho_{jm}(\omega)$ are defined as:
\begin{equation}
\rho_{jm}(\omega<\mu)=-\frac{1}{\pi L}\sum_{i}Im[\langle \psi_{0}|a_{i,jm}^{\dagger}\frac{1}{\omega-H+E_{g}+i\eta}a_{i,jm}|\psi_{0}\rangle],
\end{equation}
\begin{equation}
\rho_{jm}(\omega>\mu)=-\frac{1}{\pi L}\sum_{i}Im[\langle \psi_{0}|a_{i,jm}\frac{1}{\omega+H+E_{g}+i\eta}a_{i,jm}^{\dagger}|\psi_{0}\rangle],
\end{equation}
where $i$ is the site index. In the approach above, the sum over all the sites has to be done, which means that for every $\omega$ an ``$L$" no. of runs 
has to be performed for each site $i$ with different correction vectors $|x(\omega)\rangle = \frac{1}{\omega +i\eta + E_{g} + H}a_{i,jm}^{\dagger}|\psi_{0}\rangle$, 
and similarly for the electron part of $\rho(\omega)$. 
To reduce the cost by $``L"$ times we perform the simulations only with respect to the central site i.e. calculating the following:
\begin{equation}
\rho_{jm}(\omega, c)=-\frac{1}{\pi}Im[\langle \psi_{0}|a_{c,jm}^{\dagger}\frac{1}{\omega\pm H+E_{g}+i\eta}a_{c,jm}|\psi_{0}\rangle],
\end{equation}
where $c$ is a central site. This becomes exact for a system with periodic boundary conditions i.e. $\rho_{jm}(\omega, c)=\rho_{jm}(\omega)$, 
but our DMRG simulations are done for $L=16$ sites with open boundary conditions, for which  calculating $\rho_{jm}(\omega, c)$ is still a good approximation. 

To reproduce the data shown in this publication, the open source DMRG++ program and input files are available at {\it https://g1257.github.io/dmrgPlusPlus/} .
\newpage
\FloatBarrier
